%% file: eprint_LHCP2014_Tonjes.tex
\documentclass[10pt]{article}
\usepackage{graphicx}
\usepackage{hyperref}

\def\Title#1{\begin{center} {\Large #1 } \end{center}}
\def\Author#1{\begin{center}{ \sc #1} \end{center}}
\def\Address#1{\begin{center}{ \it #1} \end{center}}

\newcommand\pubblock{\rightline{\begin{tabular}{l} Proceedings of the Second Annual LHCP\\ \pubnumber\\
         \pubdate  \end{tabular}}}

\newenvironment{Abstract}{\begin{quotation} \begin{center} 
             \large ABSTRACT \end{center}\bigskip 
      \begin{center}\begin{large}}{\end{large}\end{center} \end{quotation}}

\newenvironment{Presented}{\begin{quotation} \begin{center} 
             PRESENTED AT\end{center}\bigskip 
      \begin{center}\begin{large}}{\end{large}\end{center} \end{quotation}}

\input econfmacros.tex

\usepackage{color}

\newcommand\pubnumber{ CMS-CR-2014-187 }

\newcommand\pubdate{\today}

\def\affiliation{
On behalf of the CMS Collaboration, \\
Department of Chemistry and Biochemistry \\
University of Maryland, College Park, MD 20742, U.S.A }

\begin{document}

\large
\begin{titlepage}
\pubblock

\vfill
\Title{Latest CMS Heavy-Ion Results on Jets}
\vfill

\Author{Marguerite Belt Tonjes}
\Address{\affiliation}
\vfill
\begin{Abstract}
Jet studies provide an experimental method to explore the features of energy loss in the strongly interacting quark-gluon plasma. Recent jet results from 2.76 TeV PbPb and pp collisions measured with the CMS detector are presented. Jets in the most head-on (central) PbPb collisions are quenched in comparison to pp, and the jets fragment in different ways in the two systems. Measurements from pPb collisions at 5.02 TeV show that the jet and charged particle suppression seen in central PbPb measurements are not due to initial state nuclear effects.
\end{Abstract}
\vfill

\begin{Presented}
The Second Annual Conference\\
 on Large Hadron Collider Physics \\
Columbia University, New York, U.S.A \\ 
June 2-7, 2014
\end{Presented}
\vfill
\end{titlepage}
\def\thefootnote{\fnsymbol{footnote}}
\setcounter{footnote}{0}
%

\normalsize 


\section{Introduction}

Heavy ions collided at high energies at the LHC provide sufficient energy density to form a strongly interacting quark-gluon plasma. The high transverse momentum partons produced by hard scatterings lose energy as they travel through this hot, dense medium~\cite{Bjorken:1982tu}. This energy loss was observed in measurements of high transverse momentum particles in central AuAu collisions at the Relativistic Heavy Ion Collider~\cite{Arsene:2004fa,Adcox:2004mh,Back:2004je,Adams:2005dq}. The high $\pt$ particles were suppressed in comparison to pp collisions taken at the same center of mass energy. The analysis was also performed with deuteron incident on Au, which showed that the significant high $\pt$ particle suppression was not due to initial state nuclear conditions. At the LHC, fully reconstructed jets are used to study the energy loss in the medium. The analyses described in these proceedings use collisions of pp at $\roots=2.76$ TeV, PbPb at $\rootsNN = 2.76$ TeV, and pPb at $\rootsNN = 5.02$ TeV measured with the CMS detector at the CERN LHC. The CMS experiment is described in reference~\cite{Chatrchyan:2008aa}.

The most central PbPb collisions are those that have the highest nuclear overlap and are represented by the lowest centrality percentiles. In central PbPb collisions, more unbalanced dijet pairs have been found than in comparison to pp~\cite{Chatrchyan:2011sx,Chatrchyan:2012nia}. In a study of the charged particle tracks relative to the leading jet axis of an unbalanced dijet pair, it was found that the imbalance inside the jet cone (of 0.8) is recovered by low $\pt$ tracks spread out over a large angle in the subleading jet direction~\cite{Chatrchyan:2011sx}. 

\section{Observations}

Jets are found utilizing the CMS particle-flow algorithm~\cite{CMS:2009nxa} which attempts to identify all stable particles in an event by combining information from all sub-detector systems. Jets are reconstructed with the anti-kT algorithm~\cite{Cacciari:2011ma} with a resolution parameter of 0.3. In inclusive pPb and PbPb jet studies, the underlying event is removed with the ``iterative-pileup" subtraction technique~\cite{Kodolova:2007hd}. For the analysis of PbPb dijet momentum flow, the underlying event is removed using the ``HF/Voronoi" technique~\cite{CMS-DP-2013-018}. The ``HF/Voronoi" technique on an event-by-event basis subtracts the underlying event based on a transverse energy azimuthal profile from the Hadron Forward (HF) detector modified using Singular Value Decomposition to take into account pseudorapidity dependent detector differences. This is followed by the equalization of calorimeter tower energies to even out non-physical negatives after underlying event subtraction.

\subsection{Nuclear Modification Factor}
\begin{figure}[!htb]
\centering
\includegraphics[height=2.2in]{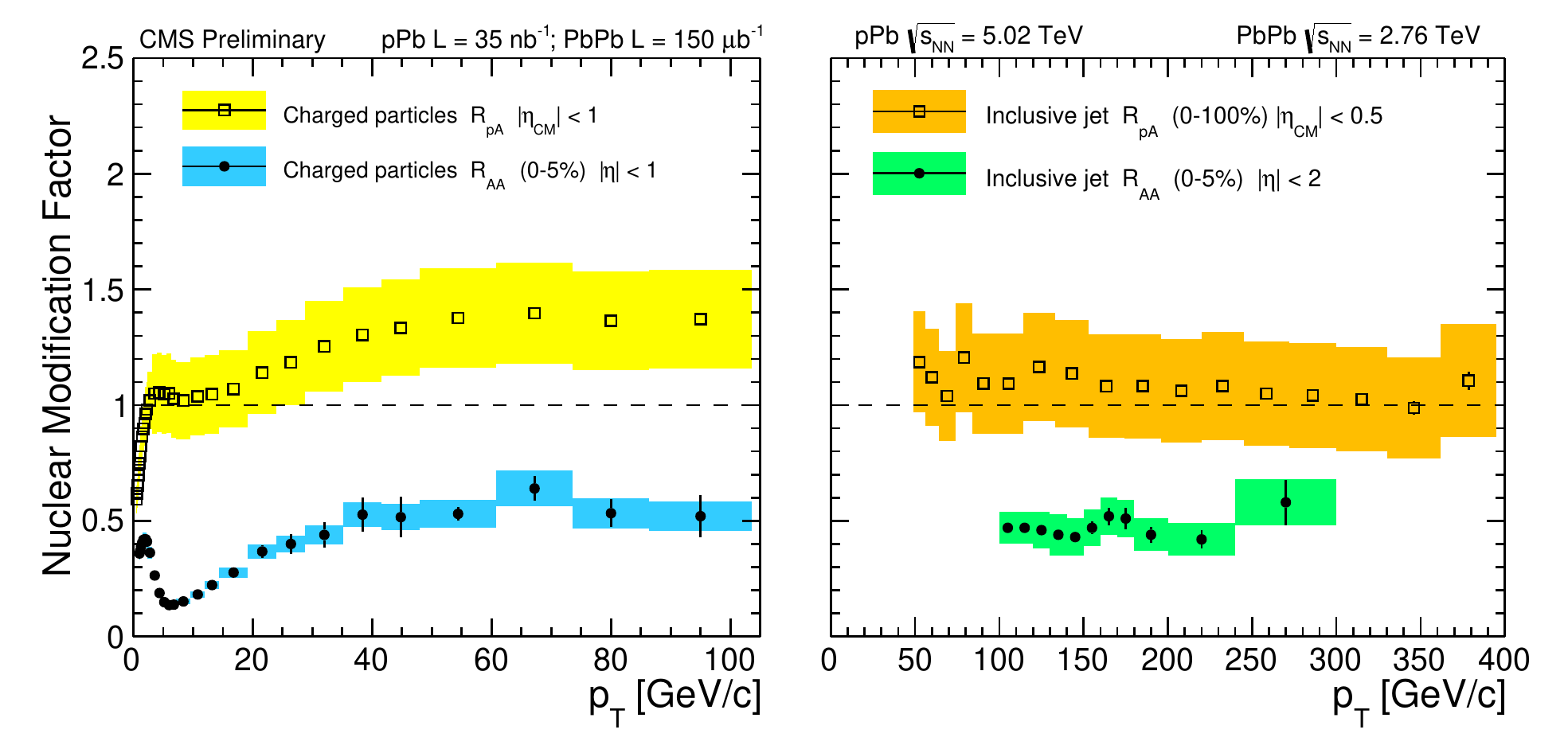}
\includegraphics[height=2.2in]{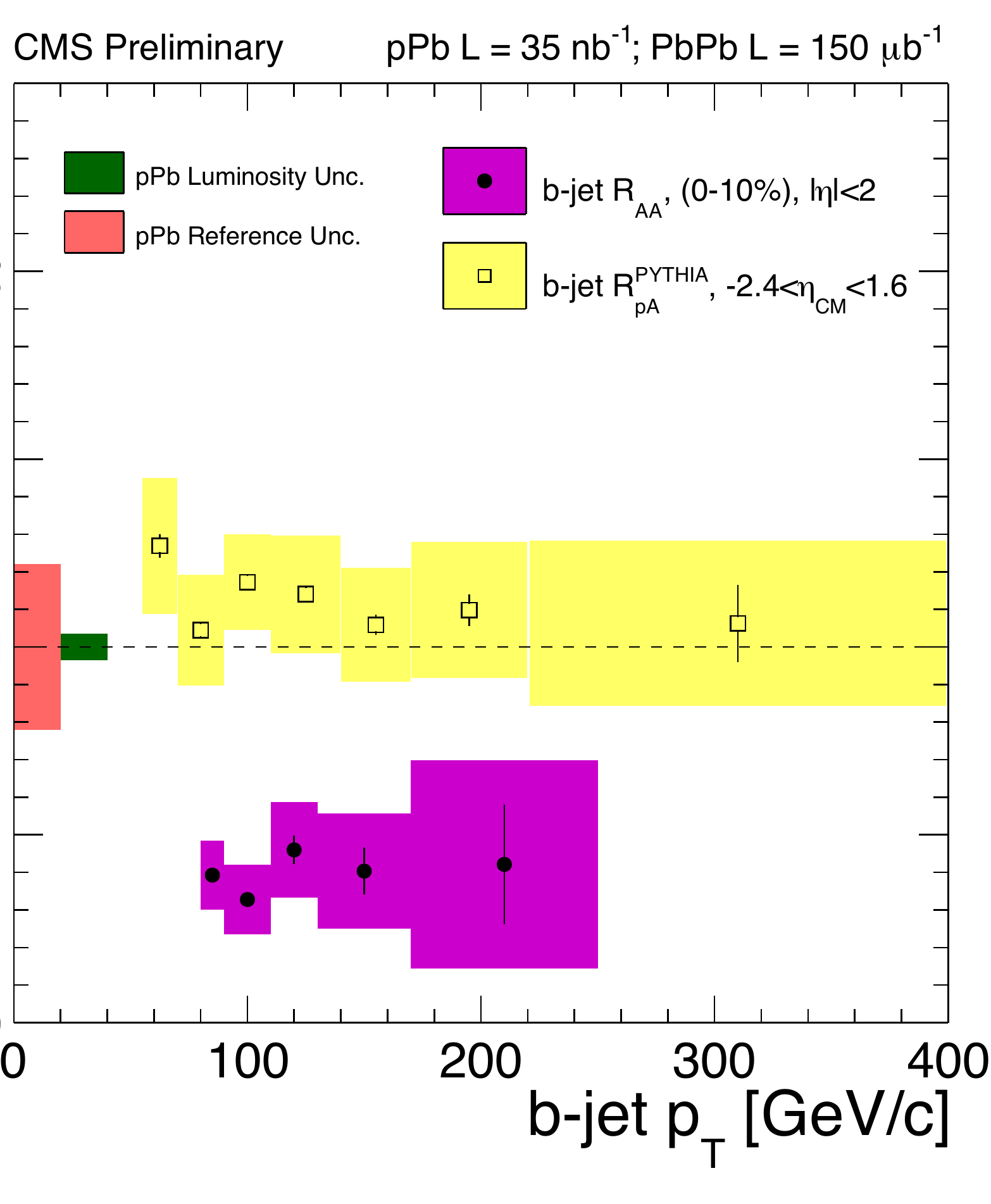}
\caption{
Left: Nuclear modification factor for charged particles measured in pPb (open squares) and central PbPb (closed circles) collisions. Center: Jet nuclear modification factor measured in pPb (open squares) and central PbPb (closed circles) collisions. Right: b-jet nuclear modification factor measured in  pPb (open squares) and central PbPb (closed circles) collisions. Statistical uncertainty is represented by bars and systematic by boxes. Luminosity uncertainty is indicated by the boxes at b-jet $p_{\mathrm{T}}=0$.
}
\label{fig:RAA}
\end{figure}

The behavior of charged particles or jets in different collisions systems is studied with the nuclear modification factor is expressed by:
\begin{center}
\begin{equation}
\label{eq:raa}
\RAA=\frac{dN^{AA}/d\pt}{\left<\ncoll\right> dN^{pp}/d\pt}=\frac{dN^{AA}/d\pt}{\left<\TAA\right>d\sigma^{pp}/d\pt},
\end{equation}
\end{center}
where $\left<\ncoll\right>$ is the average number of nucleon-nucleon collisions in heavy-ion (AA) interactions and $\left<\TAA\right>$ is the nuclear overlap function. $\left<\ncoll\right>$ is equal to  $\left<\TAA\right>~\times~\sigma^{NN}_{\mathrm{inel}}$, and is calculated with a Glauber model~\cite{Miller:2007ri} using a detailed description of the nuclear collision geometry. The nuclear modification factor shows how many charged particles or jets are measured in PbPb (or pPb) in comparison to the expectation of the average number of pp collisions superimposed. Figure~\ref{fig:RAA} (left) shows the nuclear modification factor for charged particles in pPb (open squares) in comparison to central PbPb (closed circles)~\cite{HIN-12-017-PAS,CMS:2012aa}. The pPb measurement shows a rise at low particle $\pt$ to an $\RAA$ of 1, followed by a rise at high particle $\pt$ up to 100~$\GeVc$. The reference used in the pPb measurement is an interpolation of charged particle measurements made in pp collisions at 2.76 and 7 TeV~\cite{CMS:2012aa,Chatrchyan:2011av} with CMS, as well as $|\eta|<1$ measurements at 0.63, 1.8, and 1.96 TeV from CDF~\cite{PhysRevD.82.119903,PhysRevLett.61.1819}. In contrast, the central PbPb charged particle nuclear modification factor shows a clear suppression at high particle $\pt$. Thus, the suppression observed in PbPb is not due to initial state nuclear effects.  Figure~\ref{fig:RAA} (middle) shows the nuclear modification factor for fully reconstructed jets in pPb collisions (open squares) and central PbPb collisions (closed circles)~\cite{HIN-12-004-PAS,HIN-14-001-PAS}. The pPb jets show a minimal enhancement at low jet $\pt$ in comparison to a pp reference extrapolated from CMS measurements at 7 TeV (scaled from larger jet radii)~\cite{CMS:2011ab,Chatrchyan:2012bja}. The central PbPb jets show a similar level of suppression as the high $\pt$ charged particles. Figure~\ref{fig:RAA} (right) shows the nuclear modification factor for b-jets that are tagged using distributions of the secondary vertex displacement~\cite{Chatrchyan:2013exa,HIN-14-007-PAS} compared to a reference of PYTHIA~\cite{Sjostrand:2006za} event generated b-jets embedded into HIJING~\cite{Gyulassy:1994ew}. The behavior of the pPb b-jets (open squares) and central PbPb b-jets (closed circles) are similar to that of the inclusive jet measurements.

\begin{figure}[!h]
\centering
\includegraphics[height=2.3in]{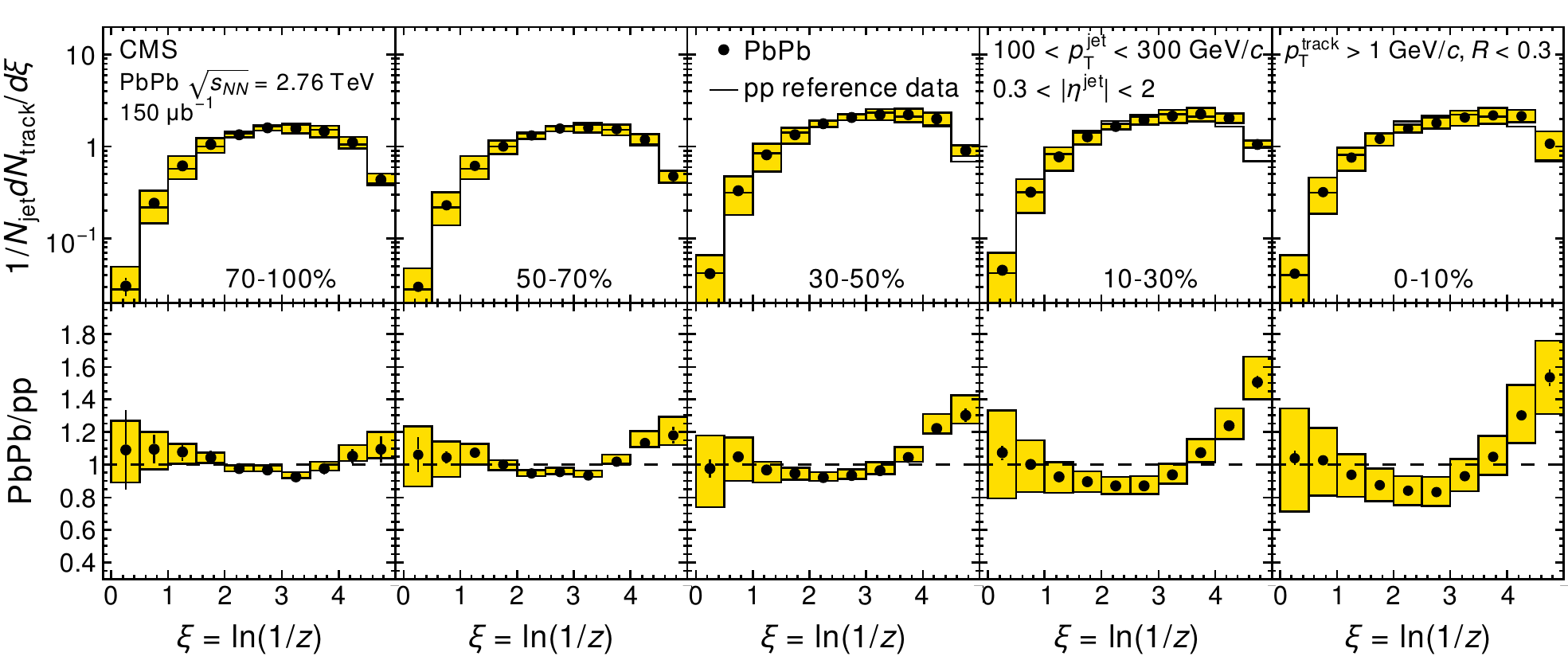}
\caption{
Top: the PbPb fragmentation function (closed circles) in bins of centrality (from peripheral on the left to central on the right) overlaid with pp reference data (horizontal lines). Jets have $100~< p_{\mathrm{T}} <~300~\GeVc$, and tracks have $p_{\mathrm{T}} > 1~\GeVc$. Bottom: the ratio of each PbPb fragmentation function to its pp reference. Error bars are statistical, and boxes show the systematic uncertainty.
}
\label{fig:ff}
\end{figure}

\subsection{Jet Fragmentation Function}
The hadronization of a jet into charged particles can be classified with a fragmentation function. The fragmentation function is defined as a function of the variable $\xi$:
\begin{equation}
\label{eq:xi}
z = \frac{p_{\parallel}^{\mathrm{track}}}{p^{\mathrm{jet}}} \; ; \; \xi = \ln \frac{1}{z} .
\end{equation}

Reconstructed inclusive jets are used with $100 < {\mathrm{jet}}~p_{\mathrm{T}} <~300~\GeVc$, and tracks of $p_{\mathrm{T}} >~1~\GeVc$ in a cone of $\mathrm{radius} < 0.3$ around the jet axis to measure the jet fragmentation function. Figure~\ref{fig:ff} (top) shows the fragmentation function for PbPb collisions (closed circles) and pp collisions (horizontal lines)~\cite{Chatrchyan:2014ava}. The bottom row shows the ratio of jet fragmentation functions for PbPb collisions from peripheral (left) to the most central (right) in comparison to the same pp measured at 2.76 TeV. For the most central collisions, inside the cone of PbPb jets, an enhancement of low $\pt$ particles is observed, as well as a suppression of intermediate $\pt$ particles.

\subsection{Momentum Flow in Unbalanced Dijets}
The momentum flow of particles in a central PbPb collision is studied by calculating the angular distribution of particles relative to the dijet axis~\cite{HIN-14-010-PAS}. Dijets in this analysis are reconstructed from Calorimeter information. A dijet pair is selected in each event where the most energetic jet is the leading jet ($p_{{\rm T},1}$), and the subleading is the next most energetic jet ($p_{{\rm T},2}$). Pairs are constrained to be back-to-back ($\Delta\phi_{\mathrm{1,2}}>\frac{5~\pi}{6}$). The most unbalanced dijet pairs are selected with an asymmetry ratio of $\rm{A}_{\mathrm{J}} > 0.22$, where the dijet asymmetry ratio is defined as
\mbox{$\rm{A}_{\rm{J}} = (p_{{\rm T},1}-p_{{\rm T},2})/(p_{{\rm T},1}+p_{{\rm T},2})$}. The direction of the dijet is defined as $\phi_{\rm Dijet} = \frac{1}{2} ( \phi_{\rm 1} + (\pi - \phi_{\rm 2} ) )$, which provides underlying event cancellation differential in position. In rings of $\Delta R = \sqrt{\Delta\phi_{\rm Trk, jet}^2 + \Delta\eta_{\rm Trk, jet}^2}$, the angular flow is studied with: 
\begin{equation}
\displaystyle{\not} p_{\mathrm{T}}^{\parallel} = 
\left( \sum_{\rm i}{ -p_{\mathrm{T}}^{\rm i}\cos{(\phi_{\rm i}-\phi_{\rm Dijet})}}\right) \Bigg|_{\mathrm{R}_{\mathrm{down}} < \Delta R < \mathrm{R}_\mathrm{up}}.
\end{equation}

\begin{figure}[!htb]
\centering
\includegraphics[height=3in]{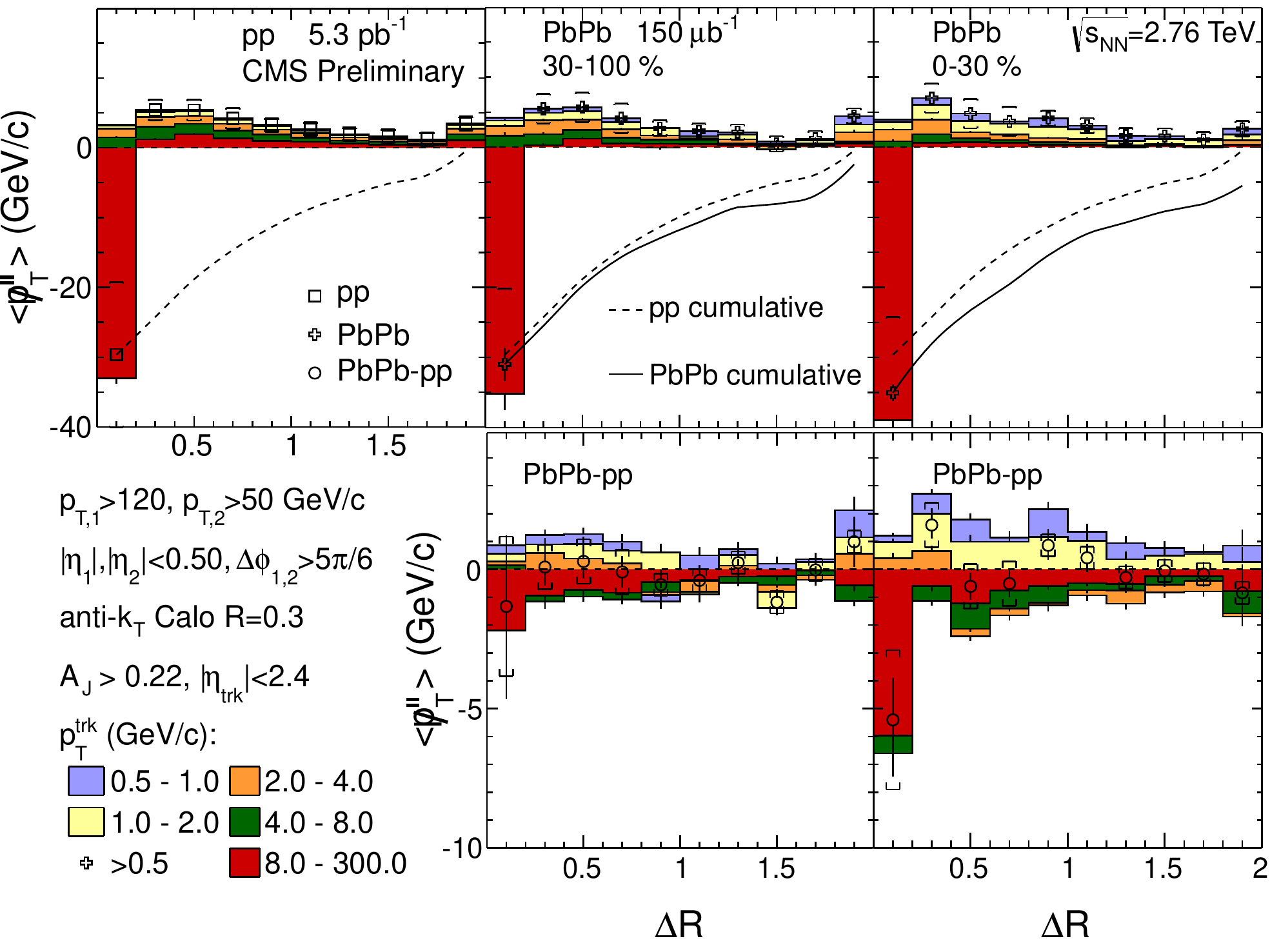}
\caption{Upper row: Differential $\displaystyle{\not} p_{\mathrm{T}}^{\parallel}$ distributions for pp, 30--100\% and 0--30\% PbPb data for various $p_{\mathrm{T}}$~ranges (colored boxes), as a function of $\Delta R$ for unbalanced dijets ($\rm{A}_{\mathrm{J}} > 0.22$). Also shown is the total $\displaystyle{\not} p_{\mathrm{T}}^{\parallel}$ (i.e., integrated over all $p_{\mathrm{T}}$ for a given $\Delta R$ bin) as a function of $\Delta R$ for pp (open squares) and PbPb data (x symbols). Dashed lines (pp) and solid lines (PbPb) show the cumulative $\displaystyle{\not} p_{\mathrm{T}}^{\parallel}$ (i.e., \ integrating the total missing $p_{\mathrm{T}}$~over $\Delta R$ starting at $\Delta R$ =0). Lower row: Difference between the PbPb and pp differential $\displaystyle{\not} p_{\mathrm{T}}^{\parallel}$ distributions per \pt range as a function of $\Delta R$ (colored boxes) and difference of the total $\displaystyle{\not} p_{\mathrm{T}}^{\parallel}$ as a function of $\Delta R$ (open circles). The bars and brackets represent statistical and systematic uncertainties respectively.
}
\label{fig:misspt}
\end{figure}

\begin{figure}[!htb]
\centering
\includegraphics[height=2.9in]{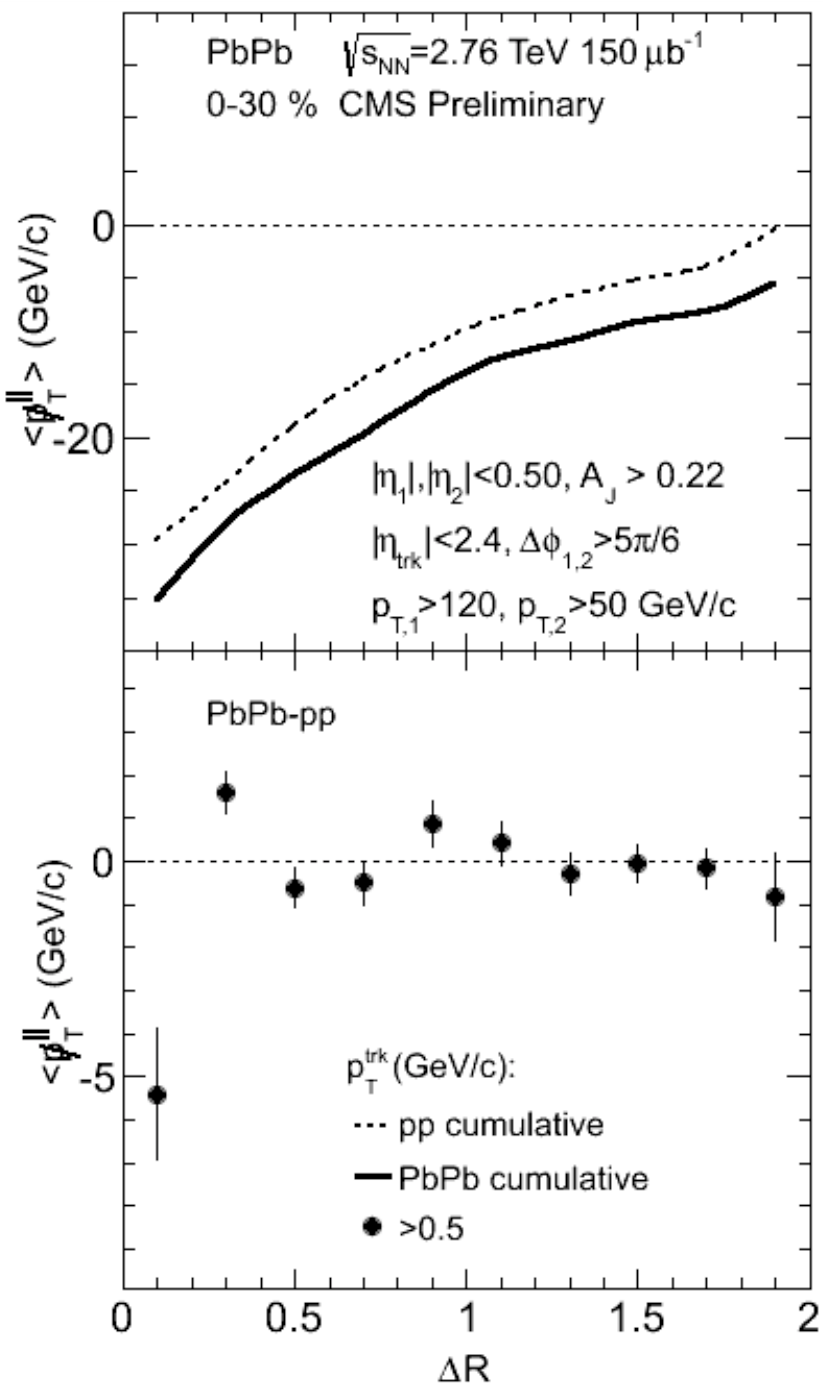}
\caption{Top: cumulative $\displaystyle{\not} p_{\mathrm{T}}^{\parallel}$ distributions for pp (dashed lines) and 0--30\% PbPb data (solid lines) as a function of $\Delta R$ for unbalanced dijets ($\rm{A}_{\mathrm{J}} > 0.22$).  Bottom: difference between the PbPb and pp cumulative $\displaystyle{\not} p_{\mathrm{T}}^{\parallel}$ distributions as a function of $\Delta R$. The bars represent statistical uncertainties.
}
\label{fig:misspt_cumulative}
\end{figure}

Figure~\ref{fig:misspt} (top) shows the analysis of charged particle track angular momentum flow measured in collisions with pp on the top left, PbPb peripheral in the middle, and PbPb central in the right for unbalanced dijets. The charged particle tracks are shown for different $\pt$ ranges, with red being the highest $\pt > 8~\GeVc$, and light blue indicating the lowest tracks of $0.5 < \pt < 1.0~\GeVc$. Focusing on central PbPb collisions at low $\Delta R$ of 0.2, it can be seen that about $35~\GeVc$ of high $\pt$ tracks are missing from the subleading jet. As the $\Delta R$ is increased, this missing $\pt$ is balanced by low $\pt$ particles up to a very large $\Delta R$ of 2.0. Figure~\ref{fig:misspt} (bottom) shows the difference between PbPb and pp collisions. The unbalanced dijets in pp collisions show evidence of a third jet with excess in high $\pt$ tracks. However, the distribution of track $\pt$ is different in the central unbalanced PbPb events. The cumulative angular distribution of the two systems is shown in Figure~\ref{fig:misspt_cumulative} (top), with the difference (PbPb-pp) shown in closed circles on the bottom. Even with very different causes for the asymmetric dijets in central PbPb and pp collisions, the total angular pattern of the momentum flow is similar within $\pm 1~\GeVc$ for $\Delta R >~0.2$.

\section{Conclusions}

Measurements of the nuclear modification factor of high $\pt$ charged particles, inclusive jets, and b-tagged jets show a similar level of strong suppression in central PbPb collisions. In comparison, the nuclear modification factor in pPb collisions does not show suppression, indicating that the central PbPb suppression is not due to initial state conditions. The charged particle $R_{\mathrm{pA}}$ shows an unexpected enhancement at high $\pt$.
Jets measured in pp and central PbPb have different fragmentation patterns, with tracks suppressed at moderate $\pt$ and enhanced at low $\pt$ for central PbPb collisions. The momentum flow of charged particles in unbalanced dijet pairs in PbPb and pp collisions was studied as a function of radius from the dijet axis. On average in central PbPb collisions, there is $35~\GeVc$ of $\pt$ missing from the most quenched jet of an asymmetric back-to-back dijet pair. The missing $\pt$ is made up of low $\pt \leq~2~\GeVc$ tracks up to $\Delta R=2.0$ around the dijet axis. Although jets in pp and central PbPb collisions fragment in different ways, the total angular pattern of momentum flow is similar, within $\pm 1~\GeVc$ for $\Delta R> 0.2$. 


\end{document}

%% file: econfmacros.tex
\def\beq{\begin{equation}}
\def\eeq#1{\label{#1}\end{equation}}
\def\eeqn{\end{equation}}

\def\beqa{\begin{eqnarray}}
\def\eeqa#1{\label{#1}\end{eqnarray}}
\def\eeqan{\end{eqnarray}}

\let\bar=\overbar

\def\Dslash{\not{\hbox{\kern-4pt $D$}}}
\def\dslash{\not{\hbox{\kern-2pt $\del$}}}

\def\msb{{\bar{\ssstyle M \kern -1pt S}}}

\newcommand{\pt}{\ensuremath{p_{\mathrm{T}}}}

\newcommand {\TAA}        {\ensuremath{T_{\rm AA}}}

\newcommand {\ncoll}    {\ensuremath{N_{\rm coll}}}
\newcommand {\RAA}   {\ensuremath{R_{\mathrm{AA}}}}
\newcommand {\rootsNN}  {\ensuremath{\sqrt{s_{_{NN}}}}}
\newcommand {\roots}    {\ensuremath{\sqrt{s}}}
\newcommand{\GeVc} {\mathrm{GeV}/\ensuremath{c}}